\documentclass[twocolumn,pre,showpacs]{revtex4}
\usepackage{amssymb}
\usepackage{amsfonts}
\usepackage{amsmath}
\usepackage{graphicx}

\begin{document}

\title{Weak ergodicity breaking induced by global memory effects}
\author{Adri\'{a}n A. Budini}
\affiliation{Consejo Nacional de Investigaciones Cient\'{\i}ficas y T\'{e}cnicas
(CONICET), Centro At\'{o}mico Bariloche, Avenida E. Bustillo Km 9.5, (8400)
Bariloche, Argentina, and Universidad Tecnol\'{o}gica Nacional (UTN-FRBA),
Fanny Newbery 111, (8400) Bariloche, Argentina}
\date{\today}

\begin{abstract}
We study the phenomenon of weak ergodicity breaking for a class of globally
correlated random walk dynamics defined over a finite set of states. The
persistence in a given state or the transition to another one depends on the
whole previous temporal history of the system. A set of waiting time
distributions, associated to each state, set the random times between
consecutive steps. Their mean value is finite for all states. The
probability density of time-averaged observables is obtained for different
memory mechanisms. This statistical object explicitly shows departures
between time and ensemble averages. While the mean residence time in each
state may result divergent, we demonstrate that this condition is in general
not necessary for breaking ergodicity. Hence, global memory effects are an
alternative mechanism able to induce this property. Analytical and numerical
calculations support these results.
\end{abstract}

\maketitle



\section{Introduction}

Ergodicity plays a fundamental role in the formulation of statistical
physics. This property is usually stated by saying that ensemble average and
time average of observables are equals, the last one being taken in the long
time (infinite) limit. In contrast with thermodynamical systems, where the
lack or ergodicity is induced by a spontaneous symmetry breaking \cite%
{goldenfeld}, the disparity between ensemble and time averages may also be
found as an emergent property of complex systems. Named as weak ergodicity
breaking (EB) \cite{bouchad}, this feature is induced by the power-law
nature of the statistical distributions associated to the observables and
their dynamics \cite{bouchad,lutz}.

Time averages in presence of weak EB remain random even in the long time
limit. Their statistics, termed as weakly non-ergodic statistical physics 
\cite{reben,barkai}, define a still very active line of research. Continuous
time random walk characterized by divergent trapping times is a natural
frame where weak EB was studied \cite{reben,barkai,bel,saa,radons,dentz}. In
addition, diverse kinds of complex anomalous diffusion processes are a
natural partner of weak EB. Analysis were performed for particles embedded
in heterogenous media \cite{cherstvy}, periodic potentials \cite{widera},
and in homogeneous disordered media \cite{garcia}. Geometric \cite{peters},
escaled \cite{safdari}\ and ultraslow \cite{bodrova} Brownian motions, as
well as diffusion induced by the combined action of different driven noises 
\cite{godec,gbel,igor}, convoluted memory processes \cite{fuli}\ and
Langevin dynamics \cite{kessler} also were characterized from a similar
perspective.

In addition to its theoretical interest, weak EB was also found in different
physical systems such as deterministic dynamics \cite%
{golan,albers,filho,akimoto} and blinking nanocrystals \cite%
{nanocrystal,margolin}; also in molecular transport \cite{manzo} and
tracking of biological single molecules \cite{burovS,burov,jeon,unkel,simon}%
\ such as lipid granules \cite{unkel}, and diffusion in the plasma membrane
of living cells \cite{simon}. Weak EB also arises in complex networks \cite%
{geneston,west}, fluid turbulence \cite{turbulence} and brain dynamics \cite%
{ross}.

Weak EB can be studied in systems that have associated a stationary state,
such as for example random walks on finite domains, and also in
non-stationary systems such as unbounded diffusive ones (see for example
Refs. \cite{reben} and \cite{burovS} respectively). Independently of the
dimensionality, weak EB is in general associated or related to some
underlying self-similiar (effective) mechanism characterized by power-law
distributions. The main goal of this paper is to demonstrate that systems
whose dynamics involves global memory effects may also develop EB.
Furthermore, we establishes that the lack of ergodicity may happens even in
absence of statistical properties (residence times) characterized by
dominant power-law distributions.

Global memory (or correlation) effects refer to systems whose stochastic
dynamics at a given time depends on its whole previous temporal history
(trajectory). These kinds of dynamics has been studied previously \cite%
{gunter,gandi,kenkre,katja,boyer,kursten,kim,esguerra,hanel}, mainly as a
mechanism that induces superdiffusion. In contrast, here we study random
walk processes defined over a finite set of states where the persistence in
a given state or the transition to another one depends on the previous
system trajectory. The random times between consecutive steps is defined by
a set of waiting time distributions with finite average times. In addition,
our main results rely on alternative memory mechanisms. They are related to
a P\'{o}lya urn dynamics \cite{feller,norman,pitman,queen,budini}, which is
one of the simplest models of contagion process, being of interest in
various disciplines \cite{norman}. In contrast to other global correlation
mechanisms, the urn-like dynamics is able to induce weak EB. Interestingly,
the departure from ergodicity arises even when the (average) residence times
in each state are finite.

The paper is organized as follows. In Sec. II, we introduce the globally
correlated random walk model. The probability density of time-averaged
observables is obtained in general. In Sec. III, we study three different
global memory mechanisms: the elephant random walk model, a random walk
driven by an urn-like dynamics, and an imperfect case of the last one. In
Sec. IV, for all models, we obtain the probability density of the residence
times. Sec. V is devoted to the Conclusions. Analytical calculations that
support the main results are presented in the Appendixes.

\section{Finite random walk with global memory effects}

In this section we introduce the globally correlated random walk model and
study its properties. The probability density of time-averaged observables
is also obtained.

\subsection{Model}

The system is characterized by a finite set of states $\mu =1,\cdots L.$ To
each state $\mu $ we assign a waiting time distribution $w_{\mu }(t),$ which
gives the statistics of times between consecutive steps of the stochastic
dynamics. We assume that all average times%
\begin{equation}
\tau _{\mu }\equiv \int_{0}^{\infty }dtw_{\mu }(t)t,  \label{TauMedio}
\end{equation}%
are finite, $\tau _{\mu }<\infty .$

The stochastic dynamics is as follows. At the beginning (initial time), each
state is selected in agreement with a set of probabilities $\{p_{\mu
}\}_{\mu =1}^{L},$ $0\leq p_{\mu }\leq 1,$ normalized as $\sum_{\mu
=1}^{L}p_{\mu }=1.$ Given that a state $\mu $ is selected, the system
remains in it during a random time selected in agreement with the waiting
time distribution $w_{\mu }(t).$ After this step, the system may remain in
the same state or jump to another one. Hence, it may \textit{persists} in
the same state, remaining an extra time interval chosen in agreement with
the same waiting time distribution, or \textit{jump} to a different state
with a different waiting time distribution. This dynamic repeats itself in
time after each step, where step refers to the process of selecting the next
state.

The state corresponding to the next step is chosen in agreement with a
conditional probability $\mathcal{T}_{n}(\{n_{1},n_{2}\cdots n_{L}\}|\mu )$
[denoted as $\mathcal{T}_{n}(\{n_{\nu }\}|\mu )].$ Here, $n$ indicates the
number of steps performed up to the present time, while $n_{\nu }$ gives the
number of times that each state $\nu $ was chosen previously. Then, $%
n=\sum_{\nu =1}^{L}n_{\nu }.$\ The dependence of the process on the whole
previous trajectory (global correlation) is given by the dependence of $%
\mathcal{T}_{n}(\{n_{\nu }\}|\mu )$ on the set $\{n_{\nu }\}_{\nu =1}^{L}.$
The previous definitions completely characterize the stochastic dynamics in
terms of the initial probabilities $\{p_{\mu }\}_{\mu =1}^{L},$ the waiting
time distributions $\{w_{\mu }(t)\}_{\mu =1}^{L}$ and the conditional (or
transition) probabilities $\mathcal{T}_{n}(\{n_{\nu }\}|\mu ).$

For the studied models [see Eqs. (\ref{BAmodel}), (\ref{PolyaModel}), and (%
\ref{MixedModel})], as a consequence of the memory effects, the following
property is observed. In the long time limit $(t\rightarrow \infty ),$ which
also correspond to a divergent number of steps $(n\rightarrow \infty ),$ the
fractions%
\begin{equation}
f_{\mu }=\lim_{n\rightarrow \infty }\frac{n_{\mu }}{n},  \label{fracciones}
\end{equation}%
$\sum_{\mu =1}^{L}f_{\mu }=1,$ may become random variables whose values
depend on each particular realization. Their probability density is denoted
by $\mathcal{P}(\{f_{\mu }\}),$ which satisfies the normalization condition $%
\int_{\Lambda }df_{1}\cdots df_{L-1}\mathcal{P}(\{f_{\mu }\})=1.$ Here, $%
\Lambda $ is the region defined by the condition $\sum_{\nu =1}^{L}f_{\mu
}=1.$ The average of $f_{\mu }$ over an ensemble realizations, denoted by $%
\langle \cdots \rangle ,$ is%
\begin{equation}
\langle f_{\mu }\rangle =\int_{\Lambda }df_{1}\cdots df_{L-1}\ f_{\mu }%
\mathcal{P}(\{f_{\nu }\}).  \label{qu}
\end{equation}

At a given time $t,$ with $P_{\mu }(t)$ we denote the (ensemble) probability 
$[\sum_{\mu =1}^{L}P_{\mu }(t)=1]$ that the system is in the (arbitrary)
state $\mu .$ This object is characterized in Appendix \ref{stationary} from
the dynamics defined previously. The stationary probability reads $P_{\mu }^{%
\mathrm{st}}\equiv \lim_{t\rightarrow \infty }P_{\mu }(t).$\ It can be
written in terms of $\mathcal{P}(\{f_{\nu }\})$ as%
\begin{equation}
P_{\mu }^{\mathrm{st}}=\left\langle \frac{f_{\mu }\tau _{\mu }}{\sum_{\mu
^{\prime }=1}^{L}f_{\mu ^{\prime }}\tau _{\mu ^{\prime }}}\right\rangle ,
\label{PEqui}
\end{equation}%
where $\tau _{\mu }$ is defined by Eq. (\ref{TauMedio}). In Appendix \ref%
{stationary} we also derive this result. Basically it say us that in each
realization the system reaches a (random) stationary state defined by the
weights $(f_{\mu }\tau _{\mu })/\sum_{\mu ^{\prime }=1}^{L}f_{\mu ^{\prime
}}\tau _{\mu ^{\prime }}.$ In consequence, $P_{\mu }^{\mathrm{st}}$ depends
on which memory mechanism drives the stochastic dynamics.

\subsection{Time-averaged observables}

To each state $\mu ,$ we assign an observable with value $\mathcal{O}_{\mu
}. $ Hence, each realization of the random walk defines a corresponding
trajectory $\mathcal{O}(t).$ In the stationary regime, its \textit{ensemble
average} $\langle \mathcal{O}\rangle _{\mathrm{st}}\equiv \lim_{t\rightarrow
\infty }\langle \mathcal{O}(t)\rangle =\lim_{t\rightarrow \infty
}\sum\nolimits_{\mu =1}^{L}P_{\mu }(t)\mathcal{O}_{\mu },$ is%
\begin{equation}
\langle \mathcal{O}\rangle _{\mathrm{st}}=\sum\nolimits_{\mu =1}^{L}P_{\mu
}^{\mathrm{st}}\mathcal{O}_{\mu },  \label{OEstacion}
\end{equation}%
where the weights follows from Eq. (\ref{PEqui}). On the other hand, its 
\textit{time average} is defined as $\mathcal{O}\equiv \lim_{t\rightarrow
\infty }(1/t)\int_{0}^{t}dt^{\prime }\mathcal{O}(t^{\prime }),$ which leads
to%
\begin{equation}
\mathcal{O}=\lim_{t\rightarrow \infty }\sum_{\mu =1}^{L}\Big{(}\frac{t_{u}}{t%
}\Big{)}\mathcal{O}_{\mu }.  \label{Observable}
\end{equation}%
Here, $t_{u}$\ is the total residence time in the state $\mu $ in the
interval $(0,t).$ Hence, $\sum\nolimits_{\mu =1}^{L}t_{u}=t.$

Even when a long time limit is present in the previous definition, the
observable $\mathcal{O}$ may be a random object that depends on each
particular realization. Its probability density can be written as%
\begin{equation}
P(\mathcal{O})=\lim_{t\rightarrow \infty }\left\langle \delta \Big{(}%
\mathcal{O}-\sum\nolimits_{\mu =1}^{L}\frac{t_{u}}{t}\mathcal{O}_{\mu }%
\Big{)}\right\rangle ,  \label{P(O)}
\end{equation}%
where, as before, $\left\langle \cdots \right\rangle $ denotes average over
an ensemble of realizations and $\delta (x)$ is the Dirac delta function.
Now, our goal is to calculate this object for the dynamics defined
previously.

Given that the waiting time distributions are characterized by a finite
average time $\tau _{\mu },$ Eq. (\ref{TauMedio}), after invoking the law of
large numbers, in the long time limit the total residence time $t_{u}$ in
each state can be approximated as $t_{u}\simeq n_{\mu }\tau _{\mu }.$
Consistently, the present time is $t\simeq \sum\nolimits_{\mu =1}^{L}n_{\mu
}\tau _{\mu }.$ Hence, we can write%
\begin{equation}
\lim_{t\rightarrow \infty }\frac{t_{u}}{t}\simeq \lim_{n\rightarrow \infty }%
\frac{n_{\mu }\tau _{\mu }}{\sum\nolimits_{\mu ^{\prime }=1}^{L}n_{\mu
^{\prime }}\tau _{\mu ^{\prime }}}=\frac{f_{\mu }\tau _{\mu }}{%
\sum\nolimits_{\mu ^{\prime }=1}^{L}f_{\mu ^{\prime }}\tau _{\mu ^{\prime }}}%
,
\end{equation}%
where the last relation follows from Eq. (\ref{fracciones}). Taking into
account that the fractions $\{f_{\mu }\}_{\mu =1}^{L}$ are characterized by
the distribution $\mathcal{P}(\{f_{\mu }\}),$ Eq.~(\ref{P(O)}) becomes%
\begin{eqnarray}
P(\mathcal{O}) &=&\int_{\Lambda }df_{1}\cdots df_{L-1}\mathcal{P}(\{f_{\mu
}\})  \notag \\
&&\times \delta \Big{(}\mathcal{O}-\sum\nolimits_{\mu =1}^{L}\frac{f_{\mu
}\tau _{\mu }}{\sum\nolimits_{\mu ^{\prime }=1}^{L}f_{\mu ^{\prime }}\tau
_{\mu ^{\prime }}}\mathcal{O}_{\mu }\Big{)}.  \label{Distribution}
\end{eqnarray}%
Therefore, $P(\mathcal{O})$ can be completely characterized after knowing
the distribution $\mathcal{P}(\{f_{\mu }\}).$ Notice that the specific
structure of the waiting time distributions $\{w_{\mu }(t)\}_{\mu =1}^{L}$
only appears through the average times $\{\tau _{\mu }\}_{\mu =1}^{L},$ Eq. (%
\ref{TauMedio}).

\subsection{Ergodicity and localization}

For an ergodic dynamics the fractions $f_{\mu }$ [Eq. (\ref{fracciones})]
must be characterized by their ensemble average, Eq. (\ref{qu}). Hence,%
\begin{equation}
\mathcal{P}(\{f_{\mu }\})=\delta (f_{1}-\langle f_{1}\rangle )\delta
(f_{2}-\langle f_{2}\rangle )\cdots \delta (f_{L}-\langle f_{L}\rangle ).
\label{Delta}
\end{equation}%
Inserting this expression into Eq. (\ref{Distribution}), it follows the
distribution%
\begin{equation}
P(\mathcal{O})=\delta (\mathcal{O-}\langle \mathcal{O}\rangle _{\mathrm{st}%
}),  \label{DeltaErgodico}
\end{equation}%
where $\langle \mathcal{O}\rangle _{\mathrm{st}}$\ is given by Eq. (\ref%
{OEstacion}) with the weights%
\begin{equation}
P_{\mu }^{\mathrm{st}}=\frac{\langle f_{\mu }\rangle \tau _{\mu }}{\sum_{\mu
^{\prime }=1}^{L}\langle f_{\mu ^{\prime }}\rangle \tau _{\mu ^{\prime }}}.
\end{equation}%
From Eqs. (\ref{PEqui}) and (\ref{Delta}), we note that these weights
correspond to the stationary probabilities of each state $\mu $ in the
ergodic case. Hence, time averages and ensemble averages do in fact coincide.

The maximal departure with respect to ergodicity happens when the dynamics
localize, that is, the system remains in the initial condition. This case
corresponds to%
\begin{equation}
\mathcal{P}(\{f_{\mu }\})=\sum_{\mu =1}^{L}p_{\mu }\delta (f_{1})\cdots
\delta (f_{\mu }-1)\cdots \delta (f_{L}).  \label{Localizado}
\end{equation}%
Hence, Eq. (\ref{Distribution}) becomes%
\begin{equation}
P(\mathcal{O})=\sum_{\mu =1}^{L}p_{\mu }\delta (\mathcal{O}-\mathcal{O}_{\mu
}).  \label{PLocalization}
\end{equation}%
These limits are reached by the following memory mechanisms.

\section{Examples}

In the examples worked below, the stochastic dynamics may reach both the
ergodic and localized regimes Eqs. (\ref{DeltaErgodico}) and (\ref%
{PLocalization}) respectively. The distribution $\mathcal{P}(\{f_{\mu }\})$
can be explicitly calculated and then the non-ergodic properties
characterized through Eq. (\ref{Distribution}).

\subsection{Elephant random walk model}

This correlation model has been studied extensively in the recent literature
as a mechanism for inducing superdiffusion \cite{gunter,kim,kursten}. In the
present context, it is defined by the transition probability%
\begin{equation}
\mathcal{T}_{n}(\{n_{\nu }\}|\mu )=\varepsilon q_{\mu }+(1-\varepsilon )%
\frac{n_{\mu }}{n}.  \label{BAmodel}
\end{equation}%
The positive weights $0<q_{\mu }<1$ are extra parameters normalized as $%
\sum_{\mu =1}^{L}q_{\mu }=1.$ The parameter $\varepsilon $ assumes values in
the interval $[0,1].$ The stochastic dynamics can be read as follows. With
probability $\varepsilon ,$ and independently of the previous history, the
new state is chosen in agreement with the probabilities $\{q_{\mu }\}_{\mu
=1}^{L}.$ On the other hand, with probability $(1-\varepsilon )$ each state
is chosen in agreement with the weights $\{n_{\mu }/n\}_{\mu =1}^{L},$ which
in fact depend on the whole previous history of the process.

For $\varepsilon =1,$ the selection of the new state is completely random
and independent of the previous history. Therefore, the system is ergodic in
this case, Eq. (\ref{DeltaErgodico}). On the other hand, for $\varepsilon =0$
the dynamics localize, that is, the system remains in the initial condition,
Eq. (\ref{PLocalization}).

Even when the dynamics reaches the ergodic and localized regime, for
intermediates values $0<\varepsilon <1$ the dynamics is ergodic. This
property is demonstrated in Appendix \ref{elephant}. In fact, the
distribution $\mathcal{P}(\{f_{\mu }\})$ is delta distributed, Eq. (\ref%
{Delta}), with $\langle f_{\mu }\rangle =q_{\mu }.$

\subsection{Random walk driven by an urn-like dynamics}

In the P\'{o}lya urn dynamics \cite{feller,norman} (initially) an urn
contains many balls that, for example, are characterized by $L$ different
possible colors. At each step, one determine the color of one ball taken at
random and put into the urn one extra ball of the same color. A similar
process can be defined by starting the urn with only one ball \cite%
{pitman,queen,budini} (Blackwell-MacQueen urn). Its dynamics is defined by
the following conditional probability, which is taken as the driving memory
mechanism.

For the random walk over the $\mu =1,\cdots L$ states, we take the
conditional probability \cite{pitman,queen}%
\begin{equation}
\mathcal{T}_{n}(\{n_{\nu }\}|\mu )=\frac{\lambda q_{\mu }+n_{\mu }}{%
n+\lambda }.  \label{PolyaModel}
\end{equation}%
As before, the set of parameters $\{q_{\mu }\}_{\mu =1}^{L}$ is normalized
to one. Instead, $\lambda $ is a positive free parameter. For $\lambda
\rightarrow \infty $ the dynamics loses any dependence on the previous
history achieving in consequence an ergodic regime, Eq. (\ref{DeltaErgodico}%
). On the other hand, for $\lambda =0,$ a localized regime is achieved, Eq. (%
\ref{PLocalization}). Hence, the intermediate values of $\lambda $ avoid
this regime and in consequence one can define a nontrivial dynamics starting
from $n=1.$

For arbitrary values of $\lambda ,$ the probability density of the
(asymptotic) fractions (\ref{fracciones}) is derived in Appendix \ref%
{dirichlet}. It can be written as%
\begin{equation}
\mathcal{P}(\{f_{\mu }\})=\left\{ \sum_{\nu =1}^{L}\frac{p_{\nu }}{q_{\nu }}%
f_{\nu }\right\} D(\{f_{\mu }\}|\{\lambda q_{\mu }\}),  \label{PPolya}
\end{equation}%
where $D(\{f_{\mu }\}|\{\lambda _{\mu }\})$ is a Dirichlet distribution \cite%
{pitman,queen},%
\begin{equation}
D(\{f_{\mu }\}|\{\lambda _{\mu }\})\equiv \frac{\Gamma (\lambda )}{%
\prod_{\mu ^{\prime }}\Gamma (\lambda _{\mu ^{\prime }})}\prod_{\mu }f_{\mu
}^{\lambda _{\mu }-1}.  \label{Dirichlete}
\end{equation}%
Here, $\lambda =\sum\nolimits_{\mu =1}^{L}\lambda _{\mu }.$ The (ensemble)
average fraction reads $\langle f_{\mu }\rangle =(q_{\mu }\lambda +p_{\mu
})/(\lambda +1).$ When $p_{\nu }=q_{\nu },$ due to the normalization $%
\sum_{\nu =1}^{L}f_{\nu }=1,$ the first factor in Eq. (\ref{PPolya}) does
not contribute, and $\langle f_{\mu }\rangle =q_{\mu }.$

We notice that $\mathcal{P}(\{f_{\mu }\})$ [Eq. (\ref{PPolya})] depends on
the initial conditions $\{p_{\mu }\}_{\mu =1}^{L}.$ This property arises
from the strong memory effects that drive the underlying stochastic
dynamics. Nevertheless, this dependence is not able to cancel any of the
stationary fractions. In consequence, the initial conditions are not
relevant for breaking or not ergodicity. In fact, given that $\mathcal{P}%
(\{f_{\mu }\})$ departs from Eq. (\ref{Delta}), this model leads to EB. The
distribution $P(\mathcal{O})$ [Eq. (\ref{Distribution})] can be evaluated
from Eq. (\ref{PPolya}).

As an example, we consider a \textit{two-level system}, where the observable
is defined by $\{\mathcal{O}_{\mu }\}\rightarrow (\mathcal{O}_{2},\mathcal{O}%
_{1}),$ with $\mathcal{O}_{1}\leq \mathcal{O\leq O}_{2}.$ After integration,
we get%
\begin{equation}
P(\mathcal{O})=\frac{1}{\mathcal{N}}\frac{[\omega _{2}(\mathcal{O}_{2}-%
\mathcal{O})]^{\lambda _{1}-1}[\omega _{1}(\mathcal{O}-\mathcal{O}%
_{1})]^{\lambda _{2}-1}}{[\omega _{2}(\mathcal{O}_{2}-\mathcal{O})+\omega
_{1}(\mathcal{O}-\mathcal{O}_{1})]^{\lambda _{1}+\lambda _{2}}},
\label{PtimePolya}
\end{equation}%
where for shortening the expression we introduced the parameters $\lambda
_{1}\equiv \lambda q_{1},$\ $\lambda _{2}\equiv \lambda q_{2},$ and the
weights%
\begin{equation}
\omega _{1}\equiv \frac{\tau _{1}}{\tau _{1}+\tau _{2}},\ \ \ \ \ \ \ \ \
\omega _{2}\equiv \frac{\tau _{2}}{\tau _{1}+\tau _{2}}.
\end{equation}%
Here, $\tau _{1}$ and $\tau _{2}$ are the average times corresponding to the
two waiting time distributions $w_{1}(t)$ and $w_{2}(t)$ respectively [Eq. (%
\ref{TauMedio})]. The normalization constant reads $\mathcal{N}^{-1}=(%
\mathcal{O}_{2}-\mathcal{O}_{1})\omega _{1}\omega _{2}\Gamma (\alpha
_{1}+\alpha _{2})/\Gamma (\alpha _{1})\Gamma (\alpha _{2}).$ For simplicity,
in the previous expressions we assumed the initial condition $p_{\mu
}=q_{u}.\ $The case $p_{\mu }\neq q_{u}$ can be recovered from these
expressions [see Eqs. (\ref{PPolya}) and (\ref{Dirichlete})].

The model (\ref{PolyaModel}) demonstrates that global memory effects may
lead to EB. This result has a close relation with the breakdown of the
standard central limit theorem for globally correlated random variables \cite%
{budini}. On the other hand, as shown in Sec. IV, depending on the values of 
$\lambda ,$ here EB arises because the residence times in each state may be
divergent, that is, their probability density is characterized by power-law
tails. The next modified dynamics also develops EB, but does not involve
power-law statistics.

\subsection{Imperfect urn-like Model}

Here, we consider a model that can be seen as an imperfect case of the
previous one. We consider the possibility of having random state selections
that do not depend on the previous\ system history. The transition
probability reads%
\begin{equation}
\mathcal{T}_{n}(\{n_{\nu }\}|\mu )=\varepsilon q_{\mu }+(1-\varepsilon )%
\frac{\lambda q_{\mu }+M_{\mu }}{M+\lambda }.  \label{MixedModel}
\end{equation}%
The set $\{q_{\mu }\}_{\mu =1}^{L}$ is normalized as before, $0\leq
\varepsilon \leq 1,$ and $\lambda \geq 0.$ Hence, with probability $%
\varepsilon $ each state $\mu ,$\ independently of the previous trajectory,
is chosen with weight $q_{\mu }.$ Complementarily, with probability $%
(1-\varepsilon )$ the state is chosen in agreement with the urn mechanism
Eq. (\ref{PolyaModel}). In fact, here $M_{\mu }$ is the number of times that
the state $\mu $ was chosen \textit{with} the urn dynamics. Furthermore, $M$
is the number of times that the urn mechanism was applied, $M=\sum_{\mu
=1}^{L}M_{\mu }.$ In contrast with the elephant model [Eq. (\ref{BAmodel})],
here the contribution proportional to $\varepsilon $ can be think as an
error in the application of the urn dynamics.

In order to clarify the stochastic dynamics induced by Eq. (\ref{MixedModel}%
), in Fig. 1 we plot two realizations (upper panels) for a two-level system
with $\mathcal{O}_{2}=1$ and $\mathcal{O}_{1}=-1.$ Hence, the observable
realizations switch between these two values. The waiting time distributions
are exponential ones%
\begin{equation}
w_{\mu }(t)=\gamma _{\mu }\exp [-\gamma _{\mu }t],
\label{ExponentialWaiting}
\end{equation}%
with $\mu =1,2.$ In the lower panels we plotted the conditional probability $%
\mathcal{T}_{n}(\{n_{\nu }\}|\mu )$ as a function of $n.$ For clarity, each
value is continued in the real interval $(i-1,i).$ The left panels
corresponds to $\varepsilon =0.1,$ Eq. (\ref{MixedModel}), while the right
panels to $\varepsilon =0,$ that is, Eq. (\ref{PolyaModel}). In both cases $%
\mathcal{T}_{n}(\{n_{\nu }\}|\mu )$ attains stationary values for increasing 
$n$ [Eq. (\ref{fracciones})]. Nevertheless, in the case $\varepsilon =0.1$
at random values of $n$ the conditional probability collapses to the value $%
q_{\mu }.$ This effect gives the error or imperfection\ with respect to the
case $\varepsilon =0.$ 
\begin{figure}[tbp]
\includegraphics[bb=45 810 700 1110,angle=0,width=8.6cm]{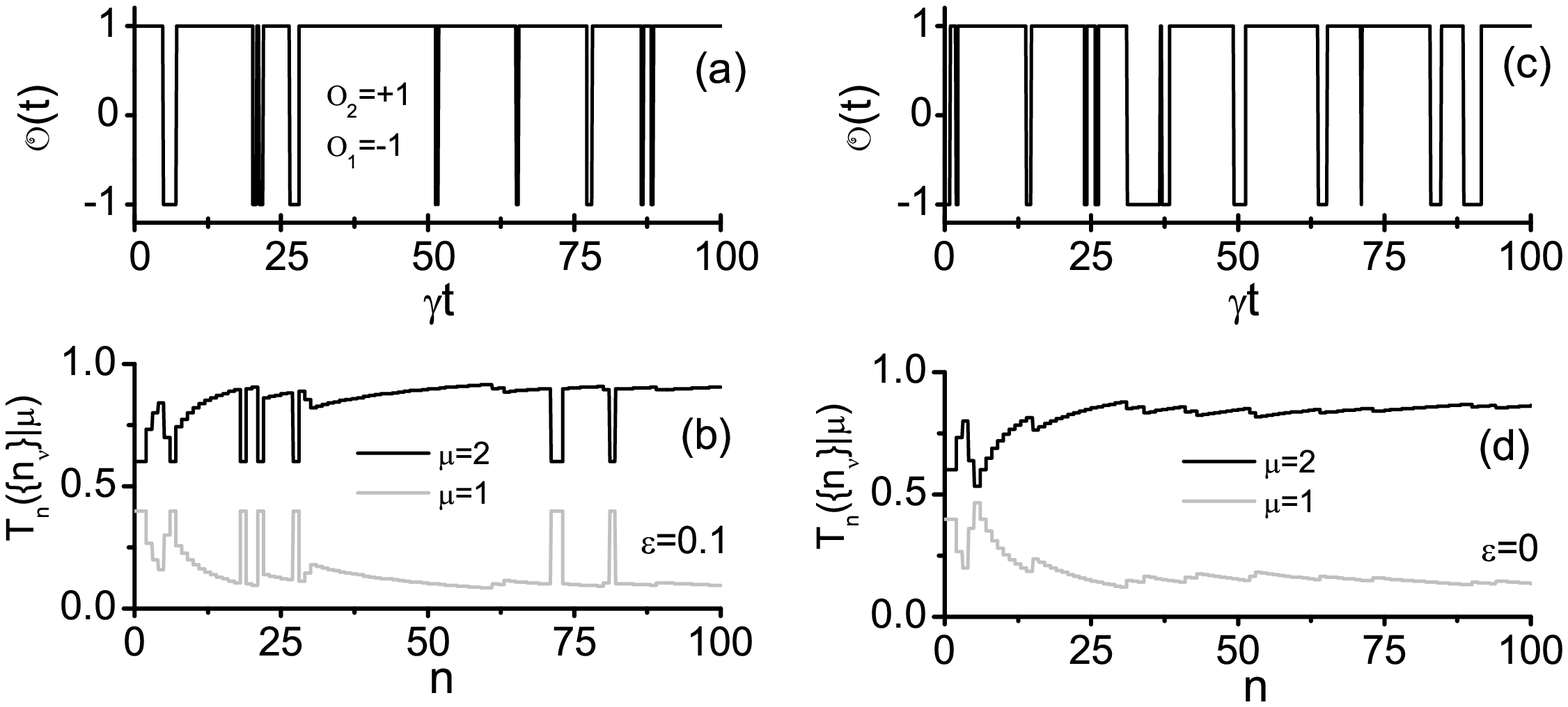}
\caption{Realizations of a two-level systems (upper panels) with observable $%
\{\mathcal{O}_{2}=1,\mathcal{O}_{1}=-1\}$ driven by an urn-like dynamics,
jointly with the corresponding conditional probabilities $\mathcal{T}%
_{n}(\{n_{\protect\nu }\}|\protect\mu )$ [Eqs. (\protect\ref{PolyaModel})
and (\protect\ref{MixedModel})] as a function of $n$ (lower panels). The
parameters are $\protect\lambda =2,$ $p_{1}=q_{1}=0.4,$ and $%
p_{2}=q_{2}=0.6. $ The waiting time distributions are exponential functions
[Eq. (\protect\ref{ExponentialWaiting})] with $\protect\gamma _{1}=\protect%
\gamma _{2}=\protect\gamma .$ In (a) and (b) $\protect\varepsilon =0.1,$
while in (c) and (d) we take $\protect\varepsilon =0.$}
\end{figure}

The probability distribution of the asymptotic fractions [Eq. (\ref%
{fracciones})] associated to Eq. (\ref{MixedModel}) is given by%
\begin{eqnarray}
\mathcal{P}(\{f_{\mu }\}) &=&\left\{ \sum_{\nu =1}^{L}\frac{p_{\nu }}{q_{\nu
}}\frac{f_{\nu }-\varepsilon q_{\nu }}{1-\varepsilon }\right\} \frac{1}{%
(1-\varepsilon )^{L-1}}  \notag \\
&&D\Big{(}\Big{\{}\frac{f_{\mu }-\varepsilon q_{\mu }}{1-\varepsilon }%
\Big{\}}|\{\lambda q_{\mu }\}\Big{)},  \label{DirReducida}
\end{eqnarray}%
where $D(\{f_{\mu }\}|\{\lambda _{\mu }\})$ is the Dirichlet distribution
Eq. (\ref{Dirichlete}). Furthermore, each fraction is restricted to the
domain%
\begin{equation}
\varepsilon q_{\mu }\leq f_{\mu }\leq 1-\varepsilon (1-q_{\mu }).
\label{DominioF}
\end{equation}%
In this case, the average fraction reads%
\begin{equation}
\langle f_{\mu }\rangle =\frac{q_{\mu }(\lambda +\varepsilon )+p_{\mu
}(1-\varepsilon )}{(\lambda +1)}.  \label{FraccionMedias}
\end{equation}

Eq. (\ref{DirReducida}) is related to Eq. (\ref{PPolya}) by the change of
variables $f_{\mu }\rightarrow \varepsilon q_{\mu }+(1-\varepsilon )f_{\mu
}. $ This relation follows by considering the asymptotic limits of Eqs. (\ref%
{MixedModel}) and (\ref{PolyaModel}), and by using that the law of large
numbers applies to the error mechanism. For $\varepsilon =0$ the previous
expressions recover the previous case, Eq. (\ref{PPolya}). Interestingly,
the effect of introducing the imperfect mechanism is to reduce the domain of
each fraction $f_{\mu }$, Eq. (\ref{DominioF}).

From Eqs. (\ref{Distribution}) and (\ref{DirReducida}) we can calculate the
distribution of the time-averaged observable. Below we consider a \textit{%
two-level system }with $\mathcal{O}_{1}<\mathcal{O}<\mathcal{O}_{2}$ and
initial condition $p_{\mu }=q_{u}.$ This case straightforwardly allows us to
reconstruct the case $p_{\mu }\neq q_{u}.$ We get%
\begin{eqnarray}
P(\mathcal{O}) &=&\frac{1}{\mathcal{N}_{\varepsilon }}[\omega _{2}(\mathcal{O%
}_{2}-\mathcal{O})-\omega _{1}^{\varepsilon }(\mathcal{O}-\mathcal{O}%
_{1})]^{\lambda _{1}-1}  \notag \\
&&\times \lbrack \omega _{1}(\mathcal{O}-\mathcal{O}_{1})-\omega
_{2}^{\varepsilon }(\mathcal{O}_{2}-\mathcal{O})]^{\lambda _{2}-1}
\label{PTimeMixto} \\
&&\times \frac{1}{[\omega _{2}(\mathcal{O}_{2}-\mathcal{O})+\omega _{1}(%
\mathcal{O}-\mathcal{O}_{1})]^{\lambda _{1}+\lambda _{2}}}.  \notag
\end{eqnarray}%
The possible values of the time-averaged observable is restricted to the
domain $\mathcal{O}_{\min }\leq \mathcal{O}\leq \mathcal{O}_{\max },$ where%
\begin{equation}
\mathcal{O}_{\max }\equiv \frac{\mathcal{O}_{2}+\mathcal{O}_{1}\omega
_{1}^{\varepsilon }\omega _{2}^{-1}}{1+\omega _{1}^{\varepsilon }\omega
_{2}^{-1}},\ \ \ \ \ \mathcal{O}_{\min }\equiv \frac{\mathcal{O}_{1}+%
\mathcal{O}_{2}\omega _{2}^{\varepsilon }\omega _{1}^{-1}}{1+\omega
_{2}^{\varepsilon }\omega _{1}^{-1}}.  \label{ObsLimits}
\end{equation}%
Furthermore, we introduced the parameters%
\begin{equation}
\omega _{1}^{\varepsilon }\equiv \omega _{1}\frac{\varepsilon q_{1}}{%
1-\varepsilon q_{1}},\ \ \ \ \ \ \ \ \omega _{2}^{\varepsilon }\equiv \omega
_{2}\frac{\varepsilon q_{2}}{1-\varepsilon q_{2}},
\end{equation}%
while the normalization constant is $\mathcal{N}_{\varepsilon }^{-1}=(%
\mathcal{O}_{2}-\mathcal{O}_{1})\omega _{1}\omega _{2}[\Gamma (\lambda
)/\Gamma (\lambda _{1})\Gamma (\lambda _{2})](1-\varepsilon )^{-(\lambda
-1)}(1-\varepsilon q_{1})^{\lambda _{1}-1}(1-\varepsilon q_{2})^{\lambda
_{2}-1}.$ Consistently, for $\varepsilon =0,$ Eq. (\ref{PTimeMixto})
recovers the previous case, Eq. (\ref{PtimePolya}). From the previous
expression it become clear that the error mechanism introduced in Eq. (\ref%
{MixedModel}) lead to a shrinking of the probability density of the
time-averaged observable. 
\begin{figure}[tbp]
\includegraphics[bb=45 580 710 1105,width=8.6cm]{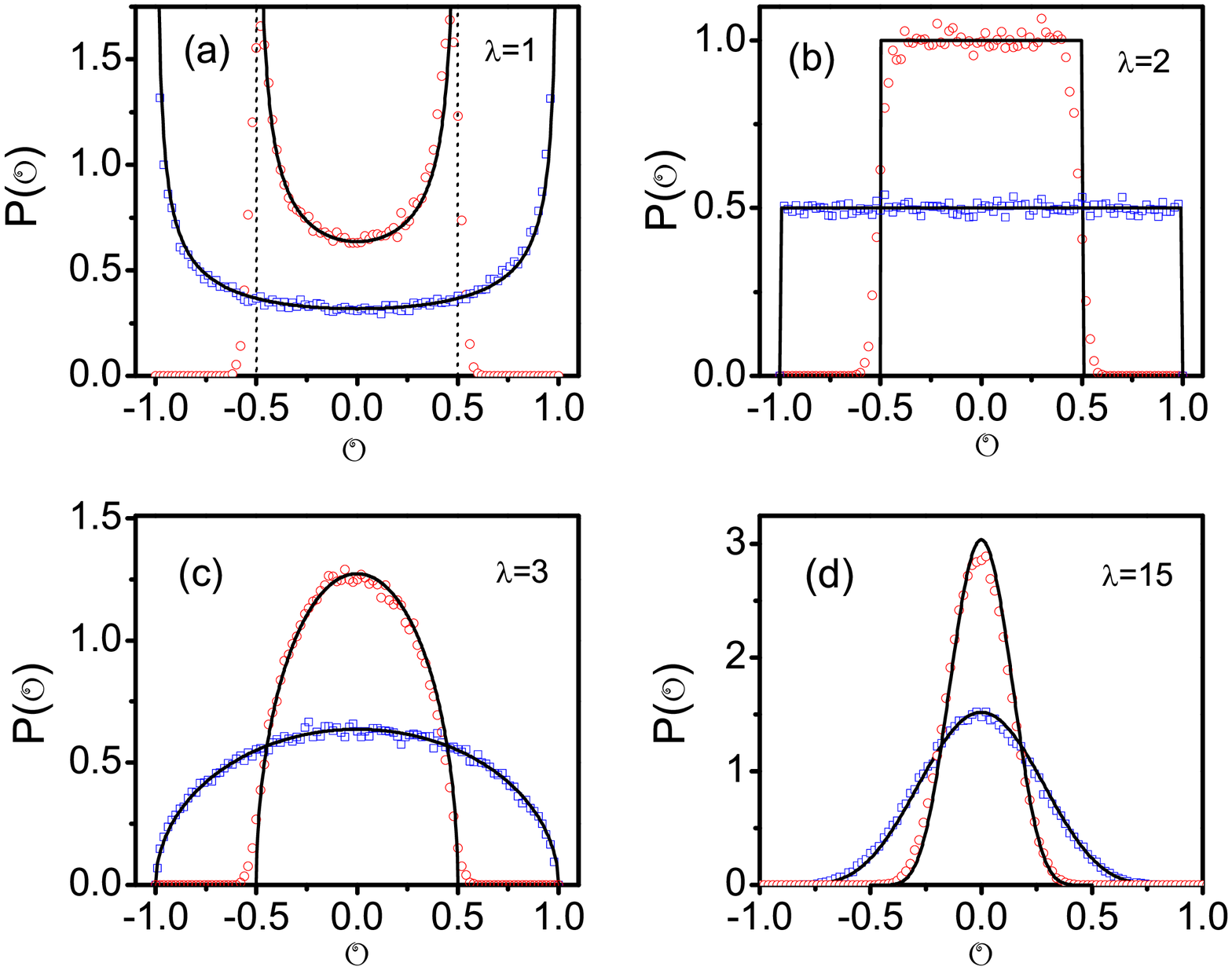}
\caption{Probability density of the time-averaged observable $\mathcal{O}$.
We take a two-level system driven by an urn-like dynamics with different
values of $\protect\lambda $ and $\protect\varepsilon .$ The full lines
correspond to the analytical expressions Eqs. (\protect\ref{PtimePolya}) and
(\protect\ref{PTimeMixto}). The waiting time distributions are exponential
functions [Eq. (\protect\ref{ExponentialWaiting})] with $\protect\gamma _{1}=%
\protect\gamma _{2}=\protect\gamma .$ In all plots we take $%
p_{1}=p_{2}=q_{1}=q_{2}=1/2.$ The (red) circles ($\protect\varepsilon =0.5$)
and (blue) squares ($\protect\varepsilon =0$) correspond to numerical
simulations. $\protect\lambda $\ is indicated in each plot.}
\end{figure}

In order to check these results, in Fig. 2, we plot the distribution (\ref%
{PTimeMixto}) for a two-level system where as before we take $\mathcal{O}%
_{2}=1,$ $\mathcal{O}_{1}=-1,$ and the exponential waiting time
distributions (\ref{ExponentialWaiting}). For each value of $\lambda ,$\ we
plot the cases $\varepsilon =0.5$ [Eqs. (\ref{MixedModel}) and (\ref%
{PTimeMixto})] and $\varepsilon =0$ [Eqs. (\ref{PolyaModel}) and (\ref%
{PtimePolya})]. Consistently, a higher $\varepsilon $ leads to a shrinking
of the density $P(\mathcal{O}),$\ which confirms that for $\varepsilon
\rightarrow 1$ an ergodic regime is achieved, $P(\mathcal{O})=\delta (%
\mathcal{O}).$ The same happen for increasing $\lambda .$ On the other hand,
the plots show that $P(\mathcal{O})$ may develops different forms such as $U$
and bell shapes, or even uniform ones. Similar dependences arise when
studying renewal random walks with divergent average trapping times \cite%
{reben}.

In all cases, the numerical simulations (circles and squares) follows from a
time average performed on a time interval with $n=10^{3}$ steps and $10^{5}$
realizations. The theoretical results fit very well the numerical ones.

\section{Probability density of residence times}

In contrast to the elephant random walk model, the previous urn models
develop weak EB. Here, we explore if this property is induced, or not, by a
power-law statistics. In fact, for continuous-time random walks with renewal
events, EB is induced by the divergence of the average residence time in
each state \cite{reben}. The residence times are the random times that the
system stays or remains in a given state before jumping to another one (see
Fig. 1). Here, for the models introduced previously, we calculate their
probability density. The calculations are valid for arbitrary number of
states $L.$

We consider a single trajectory in the \textit{long time limit}, such that
the fractions $\{f_{\mu }\}_{\mu =1}^{L}$ [Eq. (\ref{fracciones})] can be
described by their associated probability density $\mathcal{P}(\{f_{\mu }\})$
[see Eqs. (\ref{PPolya}) and (\ref{DirReducida})]. At the beginning of the
residence in a given state $\mu $ the first time interval is chosen in
agreement with its waiting time distribution $w_{\mu }(t).$ In each step,
the system remains in the same state with probability $f_{\mu },$ which add
a new random time interval also defined from $w_{\mu }(t).$ The residence
time ends when a different state $\nu \neq \mu $ is chosen. This change
occurs with probability $(1-f_{\mu }).$ Therefore, the probability $W_{\mu
}(\{f\}|\tau )d\tau $ of leaving the state $\mu $ after a residence time $%
\tau $ can be written in the Laplace domain $[g(s)=\int_{0}^{\infty }d\tau
g(\tau )e^{-s\tau }]$ as%
\begin{equation}
W_{\mu }(\{f\}|s)=(1-f_{\mu })w_{\mu }(s)\sum_{n=0}^{\infty }f_{\mu }^{n}\
w_{\mu }^{n}(s).
\end{equation}%
Here, $w_{\mu }(s)$ is the Laplace transform of the waiting time
distribution $w_{\mu }(t)$ associated to the state $\mu .$ The previous
expression takes into account all possible way of leaving the state $\mu $
after a given number of steps. It can be rewritten as%
\begin{equation}
W_{\mu }(\{f\}|s)=(1-f_{\mu })\frac{w_{\mu }(s)}{1-f_{\mu }w_{\mu }(s)}.
\label{Wu}
\end{equation}

The density\ $W_{\mu }(\{f\}|\tau )$ is a conditional object. In fact, it is
defined for a particular realization with random values of the fraction $%
f_{\mu }.$ Therefore, the probability density of the residence time $W_{\mu
}(t)$ is obtained after averaging over realizations, $W_{\mu
}(t)=\left\langle W_{\mu }(\{f\}|t)\right\rangle ,$ which is equivalent to
an average over the distribution $\mathcal{P}(\{f_{\nu }\})$ of the set of
fractions $\{f_{\nu }\}_{\nu =1}^{L}.$ Therefore, we get%
\begin{equation}
W_{\mu }(\tau )=\int_{\Lambda }df_{1}\cdots df_{L-1}\mathcal{P}(\{f_{\nu
}\})W_{\mu }(\{f\}|\tau ),  \label{WFinal}
\end{equation}%
where $W_{\mu }(\{f\}|\tau )$ follows from Eq. (\ref{Wu}) after Laplace
inversion. The average persistence time $T_{\mu }$\ is defined by%
\begin{equation}
T_{\mu }\equiv \int_{0}^{\infty }d\tau \ \tau W_{\mu }(\tau ).
\label{AverageResidence}
\end{equation}

The previous two expressions can be evaluated for arbitrary waiting time
distributions and memory models. For an exponential waiting time
distribution $w_{\mu }(t)=\gamma _{\mu }\exp [-\gamma _{\mu }t]$ [Eq. (\ref%
{ExponentialWaiting})] with mean value $\tau _{\mu }=1/\gamma _{\mu }$ [Eq. (%
\ref{TauMedio})], it follows $w_{\mu }(s)=\gamma _{\mu }/(s+\gamma _{\mu }).$
From Eq. (\ref{Wu}), we get $W_{\mu }(\{f\}|s)=(1-f_{\mu })\gamma _{\mu
}/[s+(1-f_{\mu })\gamma _{\mu }],$ which can be inverted as%
\begin{equation}
W_{\mu }(\{f\}|\tau )=(1-f_{\mu })\gamma _{\mu }\exp [-(1-f_{\mu })\gamma
_{\mu }\tau ].
\end{equation}%
For an ergodic system, characterized by the probability density $\mathcal{P}%
(\{f_{\nu }\})$ given by Eq. (\ref{Delta}), from Eq. (\ref{WFinal}) we get%
\begin{equation}
W_{\mu }(\tau )=(1-\left\langle f_{\mu }\right\rangle )\gamma _{\mu }\exp
[-(1-\left\langle f_{\mu }\right\rangle )\gamma _{\mu }\tau ].  \label{WErgo}
\end{equation}%
This result is consistent with the definition of the underlying stochastic
process that in each step allows the persistence in the same state. In fact,
the average persistence time is $T_{\mu }=1/[\gamma _{\mu }(1-\left\langle
f_{\mu }\right\rangle )],$ indicating an increasing of the average
persistence time with an increasing of the weight $\left\langle f_{\mu
}\right\rangle .$ On the other hand, in the localized regime [Eq. (\ref%
{Localizado})], due to the absence of transitions, it is not possible to
define $W_{\mu }(\tau ).$

Taking exponential waiting time distributions Eq. (\ref{ExponentialWaiting}%
), for a \textit{two-level system} $[\mu =1,2]$ characterized by the
conditional probability (\ref{MixedModel}) (imperfect urn model), after a
simple change of variables, Eqs. (\ref{DirReducida}) and (\ref{WFinal})
deliver%
\begin{equation}
W_{\mu }^{\varepsilon }(\tau )=\frac{1}{\mathcal{N}}\int_{0}^{1}df\ \varphi
_{\varepsilon }\exp [-\varphi _{\varepsilon }\ \tau ]c_{\mu }\ f^{\lambda
_{\mu }-1}(1-f)^{\lambda _{\mu ^{\prime }}-1},  \label{W_TLS}
\end{equation}%
where the super-index denotes the dependence on the parameter $\varepsilon .$
$\lambda _{\mu }=\lambda q_{\mu }$ $[\mu =1,2]$ while $\lambda _{\mu
^{\prime }}$ $[\mu ^{\prime }=2,1]$ corresponds to the other system state, $%
\lambda _{\mu ^{\prime }}=\lambda q_{\mu ^{\prime }}=\lambda (1-q_{\mu }).$
The initial conditions appears through the contribution%
\begin{equation}
c_{\mu }\equiv \frac{p_{\mu }}{q_{\mu }}f+\frac{1-p_{\mu }}{1-q_{\mu }}(1-f).
\end{equation}%
The decay rate $\varphi _{\varepsilon }$ is%
\begin{equation}
\varphi _{\varepsilon }\equiv \gamma _{\mu }[1-\varepsilon q_{\mu
}-(1-\varepsilon )f],
\end{equation}%
while the normalization constant reads $\mathcal{N}^{-1}=\Gamma (\lambda
_{1}+\lambda _{2})/\Gamma (\lambda _{1})\Gamma (\lambda _{2}).$
Straightforwardly, the average persistence time, $T_{\mu }^{\varepsilon
}=\int_{0}^{\infty }d\tau \ \tau W_{\varepsilon }(\tau ),$ from Eq. (\ref%
{W_TLS}) can then be written as%
\begin{equation}
T_{\mu }^{\varepsilon }=\frac{1}{\mathcal{N}}\int_{0}^{1}df\ \frac{1}{%
\varphi _{\varepsilon }}c_{\mu }f^{\lambda _{\mu }-1}(1-f)^{\lambda _{\mu
^{\prime }}-1}.  \label{TResidenceTLS}
\end{equation}

Consistently, for $\varepsilon =1$ Eq. (\ref{W_TLS}) recovers Eq. (\ref%
{WErgo}) with $\left\langle f_{\mu }\right\rangle =q_{\mu }$ [Eq. (\ref%
{FraccionMedias})]. Hence, $\gamma _{\mu }T_{\mu }^{1}=1/(1-q_{\mu }).$ The
same results arise when $\lambda \rightarrow \infty .$ For arbitrary $%
\varepsilon $ and $\lambda ,$ from Eqs. (\ref{W_TLS}) and (\ref%
{TResidenceTLS})\ explicit analytical expressions can be found for both $%
W_{\mu }^{\varepsilon }(\tau )$ and $T_{\mu }^{\varepsilon }$ [see Appendix %
\ref{analytical}].

Interestingly, for $0<\varepsilon \leq 1$ (and any initial condition) the
average residence time $T_{\mu }^{\varepsilon }$ is finite [see Eq. (\ref%
{TResAn})]. This is the main result of this section. In fact, this result
demonstrates that weak EB may arise even in the absence of power-law
statistical distributions with divergent average residence times. On the
other hand, for the case $\varepsilon =0,$ that is, the dynamics defined by
the conditional probabilities (\ref{PolyaModel}), the average residence time 
$T_{\mu }^{0},$ depending on the parameter values, may be finite or
infinite. From Eqs. (\ref{TResidenceTLS}) and Eq. (\ref{TResAn}) we get%
\begin{equation}
\gamma _{\mu }T_{\mu }^{0}=\frac{\lambda -\left( \frac{1-p_{\mu }}{1-q_{\mu }%
}\right) }{\lambda (1-q_{\mu })-1},\ \ \ \ \ \ \ \ \ \lambda >\frac{1}{%
(1-q_{\mu })}>1.
\end{equation}%
Consistently, for increasing $\lambda $ this expression recovers the ergodic
case [Eq. (\ref{WErgo})], $\lim_{\lambda \rightarrow \infty }\gamma _{\mu
}T_{\mu }^{0}=1/(1-q_{\mu }).$ In the complementary region of possible
values of $\lambda ,$ the average residence time is divergent,%
\begin{equation}
\gamma _{\mu }T_{\mu }^{0}=\infty ,\ \ \ \ \ \ \ \ \ \lambda \leq \frac{1}{%
(1-q_{\mu })}.  \label{infinito}
\end{equation}%
This last regime indicates that the density $W_{\mu }^{\varepsilon }(\tau )$
develops power-law tails. In fact, for long residence times, $\gamma _{\mu
}\tau \gg 1,$ from Eqs. (\ref{W_TLS}) and (\ref{ResidentGeometric}) it can
be approximated as%
\begin{equation}
W_{\mu }^{0}(\tau )\approx \gamma _{\mu }C_{\mu }^{0}\Big{(}\frac{1}{\gamma
_{\mu }\tau }\Big{)}^{\lambda (1-q_{\mu })+1},  \label{PowerAprox}
\end{equation}%
which defines the previous finite and infinite average time regimes. The
dimensionless constant reads $C_{\mu }^{0}=(p_{\mu }/q_{\mu })(1-q_{\mu
})\Gamma (1+\lambda )/\Gamma (q_{\mu }\lambda ).$ When $p_{\mu }=0$ $(p_{\mu
^{\prime }}=1)$ the asymptotic behavior becomes $W_{\mu }^{0}(\tau )\approx
(1/\gamma _{\mu }\tau )^{\lambda (1-q_{\mu })+2},$ while for $W_{\mu
^{\prime }}^{0}(\tau )$ is given by Eq. (\ref{PowerAprox}). We remark that
in general $W_{\mu }^{\varepsilon }(\tau )$ $(\varepsilon >0)$ may also
develop power-law behaviors. Nevertheless, a multiplicative exponential
factor always leads to finite averages times [see for example Eq. (\ref%
{WpartMU}) below].%
%
\begin{figure}[tbp]
\includegraphics[bb=55 26 425 556,width=7.5cm]{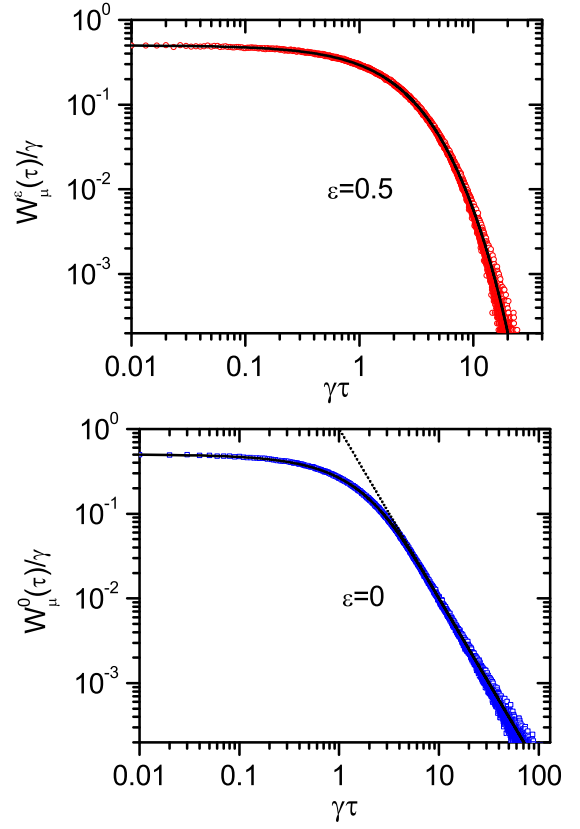} 
\caption{Probability distribution $W_{\protect\mu }^{\protect\varepsilon }(%
\protect\tau )$ $[\protect\mu =1,2]$ of the residence times for a two-level
system. The full lines correspond to the analytical result Eq. (\protect\ref%
{WpartMU}). The waiting time distributions are exponential functions [Eq. (%
\protect\ref{ExponentialWaiting})] with $\protect\gamma _{1}=\protect\gamma %
_{2}=\protect\gamma .$ In both curves, $p_{1}=p_{2}=q_{1}=q_{2}=1/2,$ and $%
\protect\lambda =2.$ The (red) circles correspond to a numerical simulation
with $\protect\varepsilon =0.5,$ while the (blue) squares to $\protect%
\varepsilon =0.$ The dotted line is the asymptotic power-law behavior (%
\protect\ref{PowerAprox}) of Eq. (\protect\ref{WCeroPart}).}
\end{figure}

For particular values of the characteristic parameters, the integral results
Eqs. (\ref{W_TLS}) and (\ref{TResidenceTLS}) lead to simple expressions.
Taking $p_{1}=q_{1}=1/2,$ $p_{2}=q_{2}=1/2,$ and $\lambda =2$ [Fig. (2b)]
the density of residence times becomes%
\begin{equation}
W_{\mu }^{\varepsilon }(\tau )=\frac{\exp (-\gamma _{\varepsilon }^{+}\tau
)(1+\gamma _{\varepsilon }^{+}\tau )-\exp (-\gamma _{\varepsilon }^{-}\tau
)(1+\gamma _{\varepsilon }^{-}\tau )}{\gamma _{\mu }\tau ^{2}(1-\varepsilon )%
},  \label{WpartMU}
\end{equation}%
where for shortening the expression we introduced the rates $\gamma
_{\varepsilon }^{+}\equiv \gamma _{\mu }\varepsilon /2$ and $\gamma
_{\varepsilon }^{-}\equiv \gamma _{\mu }(1-\varepsilon /2).$ In the case $%
\varepsilon =1$ (ergodic dynamics), we get $W_{\mu }^{\varepsilon }(\tau
)=(\gamma _{\mu }/2)T_{\mu }^{1}\exp [-(\gamma _{\mu }/2)\tau ].$ Hence, $%
T_{\mu }^{1}=2/\gamma _{\mu }.$ In the case $\varepsilon =0$ it reduces to%
\begin{equation}
W_{\mu }^{0}(\tau )=\frac{1}{\gamma _{\mu }\tau ^{2}}[1-(1+\gamma _{\mu
}\tau )\exp (-\gamma _{\mu }\tau )],  \label{WCeroPart}
\end{equation}%
which explicitly shows the presence of dominant power-law tails. The average
residence time [Eq. (\ref{TResidenceTLS})], for arbitrary $\varepsilon $
reads%
\begin{equation}
\gamma _{\mu }T_{\mu }^{\varepsilon }=\frac{2\mathrm{arctanh}(1-\varepsilon )%
}{(1-\varepsilon )}=\frac{\mathrm{\ln }\left( \frac{2-\varepsilon }{%
\varepsilon }\right) }{(1-\varepsilon )},  \label{TResParti}
\end{equation}%
where $\mathrm{arctanh}[x]=\ln \sqrt{\frac{1+x}{1-x}}$ for $x\in (-1,1).$
Thus, $T_{\mu }^{\varepsilon }$ is finite for $0<\varepsilon \leq 1.$
Consistently with Eqs. (\ref{infinito}) and (\ref{WCeroPart}), it diverges
for $\varepsilon =0,$ $T_{\mu }^{0}=\lim_{\varepsilon \rightarrow 0}T_{\mu
}^{\varepsilon }=\infty .$

In order to check the previous results we determined the distribution $%
W_{\mu }^{\varepsilon }(\tau )$ from a set of realizations such as those
shown in Fig. 1. For the same system than in Fig. 2, the results are shown
in Fig. 3. Furthermore, we take $w_{1}(t)=w_{2}(t)=\gamma \exp (-\gamma t),$
which implies $W_{1}^{\varepsilon }(\tau )=W_{2}^{\varepsilon }(\tau ).$
Consistently with the previous analytical results [Eq. (\ref{WpartMU})], for 
$\varepsilon =0.5$ [Fig. 3(a)] asymptotically the density of residence times 
$W_{\mu }^{\varepsilon }(\tau )$ is not dominated by power-law behaviors.
Instead for $\varepsilon =0$ [Fig. 3(b)] an asymptotic power-law behavior is
clearly observed [Eq. (\ref{WCeroPart})]. The numerical and theoretical
results are consistent between them.

The numerical probability densities of Fig. 3 were obtained from a set of
equally sampled realizations. This means that the same number of data for
the random residence times are taken from each realization. We took $5\times
10^{3}$ realizations with a total length of $n=5\times 10^{5}$ steps.
Furthermore, after running the dynamics during $10^{3}$ steps (long time
limit), $5\times 10^{3}$ random residence times were taken from each
realization.

\section{Summary and Conclusions}

We have introduced a random walk dynamics characterized by global memory
mechanisms. Given a finite set of states, in each step the system may remain
in the same state of jump to another one. These alternative events are
chosen from a conditional probability that depends on the whole previous
history of the system. The time between consecutive steps is determinate by
a set of waiting time distributions, all of them characterized by a finite
average time.

We focused the analysis on the ergodic properties of the stochastic
dynamics. Hence, we characterized the probability density of time-averaged
observables, [Eq. (\ref{Distribution})]. By analyzing different memory
mechanisms, we conclude that global correlations are not a sufficient
condition for breaking ergodicity, such as for example in the elephant
random walk model \ [Eq. (\ref{BAmodel})]. On the other hand, alternative
urn-like memory mechanisms [Eqs. (\ref{PolyaModel}) and (\ref{MixedModel})]
do in fact break ergodicity. In these cases, considering a two-level
dynamics, the distribution of time-averaged observables can be found in an
explicit analytical way [Eqs. (\ref{PtimePolya}) and (\ref{PTimeMixto})].

For random walks dynamics over a finite set of states, EB may be induced by
a divergent average residence time in each state. In order to cheek this
possibility for the present models, we calculated the probability density of
the residence times [Eq. (\ref{WFinal})], and the corresponding average
residence time [Eq. (\ref{AverageResidence})]. In general, the distributions
do not develop asymptotic power-law behaviors consistent with a divergent
average residence time. Hence, we conclude that global memory effects are in
fact an alternative mechanism that leads to EB. This main conclusion was
explicitly checked for two-level dynamics [Eqs. (\ref{W_TLS} ) and (\ref%
{TResidenceTLS})]. Only for a particular set of values, the residence times
have a divergent average. All previous results were confirmed by numerical
simulations [see Figs. (2) and (3)].

In conclusion, we established that weak EB may arise in systems
characterized by global memory effects. This property may emerge even when
the relevant variables are not characterized by power-law statistical
behaviors.

\section*{Acknowledgments}

This work was supported by Consejo Nacional de Investigaciones Cient\'{\i}%
ficas y T\'{e}cnicas (CONICET), Argentina.

\appendix

\section{\label{stationary}Ensemble probabilities and stationary state}

Here, we obtain the ensemble probabilities $\{P_{\mu }(t)\}_{\mu =1}^{L}$
and their corresponding long time limit, Eq. (\ref{PEqui}).

From the dynamics defined in Sec. II, the probability $P_{\mu }(t)$ that the
system is in the (arbitrary) state $\mu $ at time $t,$ can be written in the
Laplace domain $[g(s)=\int_{0}^{\infty }d\tau g(\tau )e^{-s\tau }]$ as%
\begin{eqnarray}
P_{\mu }(s) &=&P_{1}(\mu )\Phi _{\mu }(s)+\sum_{n=1}^{\infty }\sum_{\mu
_{1},\cdots \mu _{n}}P_{n+1}(\mu _{1},\cdots \mu _{n},\mu )  \notag \\
&&\times w_{\mu _{1}}(s)\cdots w_{\mu _{n}}(s)\Phi _{\mu }(s),  \label{Path}
\end{eqnarray}%
where $\Phi _{\mu }(s)=[1-w_{\mu }(s)]/s$ is the Laplace transform of the
survival probability $\Phi _{\mu }(t)=1-\int_{0}^{t}dt^{\prime }w_{\mu
}(t^{\prime }).$ Furthermore, $P_{n}(\mu _{1},\cdots ,\mu _{n})$ is the
probability of obtaining, after $n$ steps, the states $\{\mu _{1},\cdots
,\mu _{n}\}$ from the globally correlated mechanism. Hence, $P_{1}(\mu
)=p_{\mu }.$

Eq. (\ref{Path}) can be seen an addition over the ensemble realizations,
where each term gives the weight of all realizations with $n$-selection
events. Taking into account that the variables $\mu _{1},\cdots ,\mu _{n-1}$
runs over the domain of possible states $1,2,\cdots L,$ Eq. (\ref{Path}) can
also be written as%
\begin{eqnarray}
P_{\mu }(s) &=&p_{\mu }\Phi _{\mu }(s)+\sum_{n=1}^{\infty }\sum_{\{n_{\nu
}\}}P_{n}(n_{1},\cdots n_{L})  \label{PathN} \\
&&\times \mathcal{T}_{n}(\{n_{\nu }\}|\mu )\ w_{1}^{n_{1}}(s)\cdots
w_{L}^{n_{L}}(s)\Phi _{\mu }(s).  \notag
\end{eqnarray}%
Here, $P_{n}(n_{1},\cdots n_{L})$ is the joint probability of getting $%
n_{\nu }$ times the state $\nu $ after $n$-random steps, $\nu =1,\cdots L.$
Therefore, the sum $\sum_{\{n_{\nu }\}}$ is restricted to the condition $%
\sum_{\nu =1}^{L}n_{\nu }=n.$

The expression (\ref{PathN}) is exact. Now, we perform a set of
approximations for getting the stationary state $P_{\mu }^{\mathrm{st}%
}=\lim_{t\rightarrow \infty }P_{\mu }(t).$ In the long time regime, $t\gg
\{\tau _{\nu }\}_{\nu =1}^{L},$ in the Laplace domain we can approximate 
\cite{feller}\ the waiting time distribution\ as $w_{\nu }(s)\simeq 1-\tau
_{\nu }s,$ where $\tau _{\nu }$ is the average time defined by Eq. (\ref%
{TauMedio}). Therefore, $\Phi _{\mu }(s)\simeq \tau _{\mu }$ and also $%
w_{1}^{n_{1}}(s)\cdots w_{L}^{n_{L}}(s)=\prod_{\nu =1}^{L}w_{\nu }^{n_{\nu
}}(s)\simeq \exp [-s\sum_{\nu =1}^{L}\tau _{\nu }n_{\nu }],$ which in the
time domain leads to a Dirac delta function, $\delta (t-\sum_{\nu
=1}^{L}\tau _{\nu }n_{\nu }).$

In the long time regime, $n$ increases unbounded. For the studied models,
the conditional probability can then be approximated as $\mathcal{T}%
_{n}(\{n_{\nu }\}|\mu )\simeq n_{\mu }/n\simeq f_{\mu }$ [Eq. (\ref%
{fracciones})]. Consequently, Eq. (\ref{PathN}) leads to the approximation%
\begin{equation}
P_{\mu }(t)\simeq \sum_{n=1}^{\infty }\sum_{\{n_{\nu }\}}P_{n}(n_{1},\cdots
n_{L})\tau _{\mu }\frac{n_{\mu }}{n}\delta (t-\sum_{\nu =1}^{L}\tau _{\nu
}n_{\nu }).  \label{PStAprox}
\end{equation}%
By writing the delta contribution as $\delta (t-\sum_{\nu =1}^{L}\tau _{\nu
}n_{\nu })=\delta (t-n\sum_{\nu =1}^{L}\tau _{\nu }f_{\nu }),$ we realize
that in the sum over $n$ the dominant term is that with $n\simeq t/\sum_{\nu
=1}^{L}\tau _{\nu }f_{\nu }.$ Using the properties of the delta
distribution, $\delta (t-n\sum_{\nu =1}^{L}\tau _{\nu }f_{\nu
})=(1/\sum_{\nu =1}^{L}\tau _{\nu ^{\prime }}f_{\nu ^{\prime }})\delta
(n-t/\sum_{\nu =1}^{L}\tau _{\nu }f_{\nu }),$ and after the change of
variables $n_{\nu }\rightarrow f_{\nu },$ Eq. (\ref{PStAprox}) leads to the
stationary state%
\begin{equation}
P_{\mu }^{\mathrm{st}}=\int_{\Lambda }df_{1}\cdots df_{L-1}\frac{\tau _{\mu
}f_{\mu }}{\sum_{\nu =1}^{L}\tau _{\nu }f_{\nu }}\mathcal{P}(\{f_{\nu }\}),
\end{equation}%
which in fact recovers Eq. (\ref{PEqui}). This result was also checked by
numerical calculations for the memory models introduced in Sec. III.

\section{\label{elephant}Ergodicity of the elephant random walk}

The elephant random walk is defined by the transition probability (\ref%
{BAmodel}),%
\begin{equation}
\mathcal{T}_{n}(\{n_{\nu }\}|\mu )=\varepsilon q_{\mu }+(1-\varepsilon )%
\frac{n_{\mu }}{n}.
\end{equation}%
Here, we demonstrate that the fractions defined in Eq. (\ref{fracciones}), $%
f_{\mu }=\lim_{n\rightarrow \infty }(n_{\mu }/n),$ converges to $q_{\mu },$
that is, the distribution of the fractions is given by Eq. (\ref{Delta})
with $\langle f_{\mu }\rangle =q_{\mu }$ $(0<\varepsilon \leq 1).$

At a given stage, the numbers $n_{\mu }$ can be split as follows%
\begin{equation}
n_{\mu }=m_{\mu }^{(1)}+M_{\mu }^{(1)}.
\end{equation}%
Here, $m_{\mu }^{(1)}$ gives the number of times that, with probability $%
\varepsilon ,$ the state $\mu $ was chosen with probabilities $\{q_{\mu
}\}_{\mu =1}^{L}.$ Complementarily, $M_{\mu }^{(1)}$ gives the number of
times that, with probability $1-\varepsilon ,$ the state $\mu $ was chosen
with probabilities $\{n_{\mu }/n\}.$ In the limit of a diverging number of
selections (steps), the law of large numbers gives $\lim_{n\rightarrow
\infty }m_{\mu }^{(1)}/n=\varepsilon q_{\mu }.$ Thus, asymptotically we can
approximate%
\begin{equation}
\mathcal{T}_{n}(\{n_{\nu }\}|\mu )\simeq \varepsilon q_{\mu }+(1-\varepsilon
)\Big{[}\varepsilon q_{\mu }+\frac{M_{\mu }^{(1)}}{n}\Big{]}.
\end{equation}%
Now, we can split $M_{\mu }^{(1)}$ in the same way as follows%
\begin{equation}
M_{\mu }^{(1)}=m_{\mu }^{(2)}+M_{\mu }^{(2)}.
\end{equation}%
Here, $m_{\mu }^{(2)}$ is the number of times that, with probability $%
(1-\varepsilon )\varepsilon ,$ the state $\mu $\ was chosen with
probabilities $\{q_{\mu }\}_{\mu =1}^{L}.$ Similarly, $M_{\mu }^{(2)}$ gives
the number of times that, with probability $(1-\varepsilon )\times
(1-\varepsilon ),$ the state $\mu $ was chosen with probabilities $\{M_{\mu
}^{(1)}/n\}.$ By using that $\lim_{n\rightarrow \infty }m_{\mu
}^{(2)}/n=(1-\varepsilon )\varepsilon q_{\mu },$ it follows the approximation%
\begin{equation}
\mathcal{T}_{n}(\{n_{\nu }\}|\mu )\simeq \varepsilon q_{\mu }+(1-\varepsilon
)\Big{[}\varepsilon q_{\mu }+\varepsilon q_{\mu }(1-\varepsilon )+\frac{%
M_{\mu }^{(2)}}{n}\Big{]}.
\end{equation}%
Performing the same splitting, at an arbitrary order we can write%
\begin{equation}
M_{\mu }^{(k-1)}=m_{\mu }^{(k)}+M_{\mu }^{(k)},
\end{equation}%
where the law of large numbers gives $\lim_{n\rightarrow \infty }m_{\mu
}^{(k)}/n=(1-\varepsilon )^{k-1}\varepsilon q_{\mu }.$ Therefore, we get%
\begin{equation}
\mathcal{T}_{n}(\{n_{\nu }\}|\mu )\simeq \varepsilon q_{\mu }+(1-\varepsilon
)\varepsilon q_{\mu }\sum_{k=0}^{\infty }(1-\varepsilon )^{k}=q_{\mu }.
\label{terere}
\end{equation}%
This argument shows that in the asymptotic limit the memory on the previous
states is lost. Hence, the finite random walk becomes ergodic, Eq. (\ref%
{Delta}) with $\langle f_{\mu }\rangle =q_{\mu }.$ Numerical simulations
confirm this result. Notice that the previous argument does not apply to the
urn models Eqs. (\ref{PolyaModel}) and (\ref{MixedModel}). On the other
hand, we checked that for $\varepsilon \rightarrow 0$ the\ rate of
convergence to the regime defined by Eq. (\ref{terere}) is smaller, being
infinite for $\varepsilon =0,$ that is, in the localized regime. We remark
that this result does not contradict previous results for unbounded
diffusion processes \cite{gunter,kim,kursten}.

\section{\label{dirichlet}Fraction probability density of the urn-like
dynamics}

For the urn dynamics defined by Eq. (\ref{PolyaModel}), here we obtain the
probability density of the stationary fractions Eq. (\ref{fracciones}).

By using Bayes rule, the joint probability $P_{n}(\mu _{1},\cdots \mu _{n})$
of obtaining the values $\mu _{1},\cdots \mu _{n}$ with the dynamics Eq. (%
\ref{PolyaModel}) can be written as%
\begin{equation*}
P_{n}(\mu _{1},\cdots \mu _{n})\!=\!P_{1}(\mu _{1})\mathcal{T}_{1}(\{n_{\nu
_{1}}\}|\mu _{2})\cdots \mathcal{T}_{n-1}(\{n_{\nu _{n-1}}\}|\mu _{n}).
\end{equation*}%
By writing this expression in an explicit way, we realize that the joint
probability $P_{n}(n_{1},\cdots n_{L})$ of getting $n_{\mu }$ times the
state $\mu $ after $n$-random steps can be written as%
\begin{eqnarray}
P_{n}(n_{1},\cdots n_{L}) &=&\sum_{\nu =1}^{L}\frac{(n-1)!}{n_{1}!\cdots
(n_{\nu }-1)!\cdots n_{L}!}  \label{sol} \\
&&\times p_{\nu }\frac{\Gamma (\lambda )}{\Gamma (n+\lambda )}\frac{1}{%
q_{\nu }}\prod_{\mu =1}^{L}\frac{\Gamma (n_{\mu }+\lambda _{\mu })}{\Gamma
(\lambda _{\mu })},  \notag
\end{eqnarray}%
where $\lambda _{\mu }=\lambda q_{\mu }.$ Each term in the sum $\sum_{\nu
=1}^{L}$ corresponds to all realizations with the same initial condition,
which leads to the weight $p_{\nu }.$ The contributions proportional to the
Gamma functions follows straightforwardly from the product of successive
conditional probabilities $\mathcal{T}_{k}(\{n_{k}\}|\mu _{k+1})$\ and the
property $\Gamma (n+x)/\Gamma (x)=x(1+x)(2+x)\cdots (n-1+x).$ Furthermore,
in the first line the multinomial factor takes into account all realizations
with the same numbers $\{n_{\mu }\}_{\mu =1}^{L}.$ Eq. (\ref{sol}) can be
rewritten as%
\begin{equation}
P_{n}(n_{1},\cdots n_{L})=\sum_{\nu =1}^{L}\frac{p_{\nu }}{q_{\nu }}\frac{%
n_{\nu }}{n}D_{n}(n_{1},\cdots n_{L}),  \label{saturno}
\end{equation}%
where%
\begin{equation}
D_{n}(n_{1},\cdots n_{L})\equiv \frac{n!}{n_{1}!\cdots n_{L}!}\frac{\Gamma
(\lambda )}{\Gamma (n+\lambda )}\prod_{\mu =1}^{L}\frac{\Gamma (n_{\mu
}+\lambda _{\mu })}{\Gamma (\lambda _{\mu })}.  \label{salio}
\end{equation}

In the limit $x\rightarrow \infty $ it is valid the Stirling approximation $%
\Gamma (x)\approx \sqrt{2\pi /x}e^{-x}x^{x}.$ Hence, in the same limit, it
follows $\Gamma (x+\alpha )/\Gamma (x)\approx x^{\alpha }.$ Using that $%
n!=\Gamma (n+1),$ and applying the previous approximations to Eq. (\ref%
{salio}), in the limit $n\rightarrow \infty $ it follows%
\begin{equation}
D_{n}(n_{1},\cdots n_{L})\approx \frac{\Gamma (\lambda )}{n^{\lambda -1}}%
\prod_{\mu =1}^{L}\frac{n_{\mu }^{\lambda _{\mu }-1}}{\Gamma (\lambda _{\mu
})}.
\end{equation}%
By performing the change of variables $n_{\mu }\rightarrow nf_{\mu },$ and
by using that (due to normalization) there are $(L-1)$ independent variables 
$f_{\mu },$ the previous expression straightforwardly leads to the Dirichlet
distribution $D(\{f_{\mu }\}|\{\lambda _{\mu }\}),$ Eq. (\ref{Dirichlete}).
Therefore, in the same limit, Eq. (\ref{saturno}) trivially recovers Eq. (%
\ref{PPolya}).

\section{\label{analytical}Exact analytical results for two-level systems}

For two-levels systems driven by the imperfect urn dynamics, the integrals
expressions for the probability density of residence times [Eq. (\ref{W_TLS}
)] and the average residence time [Eq. (\ref{TResidenceTLS})] can be
explicitly evaluated. $W_{\mu }^{\varepsilon }(\tau )$ reads%
\begin{eqnarray}
W_{\mu }^{\varepsilon }(\tau ) &=&\gamma _{\mu }\exp [-\gamma _{\mu }\tau
(1-\varepsilon q_{\mu })]  \notag \\
&&\{a_{\mu }(\tau )\ _{1}F_{1}[\lambda q_{\mu };\lambda ;(1-\varepsilon
)\gamma _{\mu }\tau ]  \label{ResidentGeometric} \\
&&+b_{\mu }(\tau )\ _{1}F_{1}[\lambda q_{\mu };\lambda +1;(1-\varepsilon
)\gamma _{\mu }\tau ]\}.  \notag
\end{eqnarray}%
The Kummer confluent hypergeometric function is $_{1}F_{1}[a;b;z]=%
\sum_{k=0}^{\infty }(a)_{k}(b)_{k}z^{k}/k!$ with $(x)_{k}=%
\prod_{j=0}^{k-1}(x+j)=\Gamma (x+k)/\Gamma (x).$ The auxiliary function $%
a_{\mu }(\tau )$ is%
\begin{equation}
a_{\mu }(\tau )\!\equiv \!\frac{p_{\mu }}{q_{\mu }}(1-q_{\mu })\varepsilon +%
\frac{(p_{\mu }-q_{\mu })\lambda }{q_{\mu }\gamma _{\mu }\tau },
\end{equation}%
while $b_{\mu }(\tau )$ is%
\begin{equation}
b_{\mu }(\tau )\!\equiv \!(1-\frac{p_{\mu }}{q_{\mu }}\varepsilon
)-(1-\varepsilon )p_{\mu }+(p_{\mu }-q_{\mu })(\varepsilon -\frac{\lambda }{%
q_{\mu }\gamma _{\mu }\tau }).
\end{equation}

Similarly, the average residence time is%
\begin{eqnarray}
\gamma _{\mu }T_{\mu }^{\varepsilon } &=&a_{\mu }\ _{2}F_{1}[1;\lambda
q_{\mu };\lambda ;\frac{1-\varepsilon }{1-\varepsilon q_{\mu }}]
\label{TResAn} \\
&&+b_{\mu }\ _{2}F_{1}[1;1+\lambda q_{\mu };1+\lambda ;\frac{1-\varepsilon }{%
1-\varepsilon q_{\mu }}].  \notag
\end{eqnarray}%
Here, the hypergeometric function is defined by $_{2}F_{1}[a;b;c;z]=%
\sum_{k=0}^{\infty }(a)_{k}(b)_{k}(c)_{k}z^{k}/k!,$ while the coefficients
are%
\begin{equation}
a_{\mu }\equiv \frac{1-p_{\mu }}{(1-q_{\mu })(1-\varepsilon q_{\mu })},\ \ \
\ b_{\mu }\equiv \frac{p_{\mu }-q_{\mu }}{(1-q_{\mu })(1-\varepsilon q_{\mu
})}.
\end{equation}


\begin{thebibliography}{99}
\bibitem{goldenfeld} N. Goldenfeld, \textit{Lectures on phase transitions
and the renormalization group}, (Perseus, 1992).

\bibitem{bouchad} J. P. Bouchaud, Weak ergodicity breaking and aging in
disordered systems, J. Phys. I \textbf{2}, 1705 (1992).

\bibitem{lutz} E. Lutz, Power-Law Tail Distributions and Nonergodicity,
Phys. Rev. Lett. \textbf{93}, 190602 (2004).

\bibitem{reben} A. Rebenshtok and E. Barkai, Weakly Non-Ergodic Statistical
Physics, J. Stat. Phys. \textbf{133}, 565 (2008); A. Rebenshtok and E.
Barkai, Distribution of Time-averaged Observables for Weak Ergodicity
Breaking, Phys. Rev. Lett. \textbf{99}, 210601 (2007).

\bibitem{barkai} G. Margolin and E. Barkai, Nonergodisity of a time series
obeying L\'{e}vy statistics, J. Stat. Phys. \textbf{122}, 137 (2006).

\bibitem{bel} G. Bel and E. Barkai, Stochastic Ergodicity Breaking: a random
Walk Approach, Phys. Rev. Lett. \textbf{94}, 240602 (2005); G. Bel and E.
Barkai, A Random Walk to a Non-Ergodic Equilibrium Concept, Phys. Rev. E 
\textbf{73}, 016125 (2006); J. H. P. Schulz and E. Barkai, Fluctuations
around equilibrium laws in ergodic continuous-time random walks, Phys. Rev.
E \textbf{91}, 062129 (2015).

\bibitem{saa} A. Saa and R. Venegeroles, Ergodic transitions in
continuous-time random walks, Phys. Rev. E \textbf{82}, 031110 (2010).

\bibitem{radons} T. Albers and G. Radons, Subdiffusive continuous time
random walks and weak ergodicity breaking analyzed with the distribution of
generalized diffusivities, Euro Phys. Lett. \textbf{102}, 40006 (2013).

\bibitem{dentz} M. Dentz, A. Russian, and P. Gouze, Self-averaging and
ergodicity of subdiffusion in quenched random media, Phys. Rev. E \textbf{93}%
, 010101(R) (2016.)

\bibitem{cherstvy} A. G. Cherstvy, A. V. Chechkin, and R. Metzler, Anomalous
diffusion and ergodicity breaking in heterogeneous diffusion processes, New
J. of Phys. \textbf{15}, 083039 (2013); A. G. Cherstvy and R. Metzler,
Non-ergodicity, fluctuations, and criticality in heterogeneous diffusion
processes, Phys. Rev. E \textbf{90}, 012134 (2014).

\bibitem{widera} F. Kindermann, A. Dechant, M. Hohmann, T. Lausch, D. Mayer,
F. Schmidt, E. Lutz, and A. Widera, Nonergodic Diffusion of Single Atoms in
a Periodic Potential, arXiv:1601.0666 (2016); M. Khoury, A. M. Lacasta, J.
M. Sancho, and K. Lindenberg, Weak disorder: anomalous transport and
diffusion are normal yet again, Phys. Rev. Lett. \textbf{106}, 090602 (2010).

\bibitem{garcia} P. Massignan, C. Manzo, J. A. Torreno-Pina, M. F. Garc\'{\i}%
a-Parajo, M. Lewestein, and G. J. Lapeyre, Jr., Nonergodic Subdiffusion from
Brownian Motion in a Inhomogeneous Medium, Phys. Rev. Lett. \textbf{112},
150603 (2014).

\bibitem{peters} O. Peters, Ergodicity breaking in geometric Brownian
motion, Phys. Rev. Lett. \textbf{110}, 100603 (2013).

\bibitem{safdari} H. Safdari, A. G. Cherstvy, A. V. Chechkin, F. Thiel, I.
M. Sokolov, and R. Metzler, Quantifying the non-ergodicity of scaled
Brownian motion, J. Phys. A \textbf{48}, 375002 (2015); H. Safdari, A. V.
Chechkin, G. R. Jafari, and R. Metzler, Aging scaled Brownian motion, Phys.
Rev. E \textbf{91}, 042107 (2015).

\bibitem{bodrova} A. Godec, A. V. Chechkin, E. Barkai, H. Kantz, and R.
Metzler, Localization and universal fluctuations in ultraslow diffusion
processes, J. Phys. A \textbf{47}, 492002 (2014); A. S. Bodrova, A. G.
Cherstvy, A. V. Chechkin, and R. Metzler, Ultraslow scaled Brownian motion,
arXiv:1503.08125 (2015).

\bibitem{godec} A. Godec and R. Metzler, Finite-Time Effects and Ultraweak
Ergodicity Breaking in Superdiffusive Dynamics, Phys. Rev. Lett. \textbf{110}%
, 020603 (2013).

\bibitem{gbel} G. Bel and I. Nemenman, Ergodic and non-ergodic anomalous
diffusion in coupled stochastic processes, New. J. Phys. \textbf{11}, 083009
(2009).

\bibitem{igor} Y. Meroz, I. M. Sokolov and J. Klafter, Subdiffusion of mixed
origins: When ergodicity and nonergodicity coexist, Phys. Rev. E \textbf{81}%
, 010101(R) (2010); F. Thiel and I. M. Sokolov, Weak ergodicity breaking in
an anomalous diffusion process of mixed origins, Phys. Rev. E \textbf{89},
012136 (2014).\ 

\bibitem{fuli} A. Fulinski, Anomalous diffusion and weak nonergodicity,
Phys. Rev. E \textbf{83}, 061140 (2011).

\bibitem{kessler} A. Dechant, E. Lutz, D. A. Kessler, and E. Barkai,
Fluctuations of Time Averages for Langevin Dynamics in a Binding Force
Field, Phys. Rev. Lett. \textbf{107}, 240603 (2011).

\bibitem{golan} G. Bel and E. Barkai, Ergodicity Breaking in a Deterministic
Dynamical System, EuroPhys. Lett. \textbf{74}, 15 (2006);

\bibitem{albers} T. Albers and G\"{u}nter Radons, Weak Ergodicity Breaking
and Aging of Chaotic Transport in Hamiltonian Systems, Phys. Rev. Lett. 
\textbf{113}, 184101 (2014);

\bibitem{filho} A. Figueiredo, T. M. Rocha Filho, M. A. Amato, Z. T.
Oliveira, Jr., and R. Matsushita, Truncated L\'{e}vy flights and weak
ergodicity breaking in the Hamiltonian mean-field model, Phys. Rev. E 
\textbf{89}, 022106 (2014).

\bibitem{akimoto} T. Akimoto, Distributional Response to Biases in
Deterministic Superdiffusion, Phys. Rev. Lett. \textbf{108}, 164101 (2012).

\bibitem{nanocrystal} X. Brokmann, J.-P. Hermier, G. Messin, P. Desbiolles,
J.-P. Bouchaud, and M. Dahan, Statistical Aging and Nonergodicity in the
Fluorescence of Single Nanocrystals, Phys. Rev. Lett. \textbf{90}, 120601
(2003).

\bibitem{margolin} G. Margolin and E. Barkai, Nonergodicity and Blinking
Nanocrystals and Other L\'{e}vy-Walk processes, Phys. Rev. Lett. \textbf{94}%
, 080601 (2005).

\bibitem{manzo} C. Manzo, J. Arreno-Pina, P. Massignan, G. J. Lapeyre, Jr.
M. Lewenstein, and M. F. Garcia Parajo, Weak Ergodicity Breaking of Receptor
Motion in Living Cells Stemming from Random Diffusivity, Phys. Rev. X 
\textbf{5}, 011021 (2015).

\bibitem{burovS} Y. He, S. Burov, R. Metzler, and E. Barkai, Random
Time-Scale Invariant Diffusion and Transport Coefficients, Phys. Rev. Lett. 
\textbf{101}, 058101 (2008).

\bibitem{burov} A. Lubelski, I. M . Sokolov, and J. Klafter, Nonergodicity
Mimics Inhomogeneity in Single Particle Tracking, Phys. Rev. Lett. \textbf{%
100}, 250602 (2008).

\bibitem{jeon} S. Burov, J. -H. Jeon, R. Metzler, and E. Barkai, Single
particle tracking in systems showing anomalous diffusion: the role of weak
ergodicity breaking, Phys. Chem. Chem. Phys. \textbf{13}, 1800 (2011).

\bibitem{unkel} J. Jeon, V. Tejedor, S. Burov, E. Barkai, C. Selhuber-Unkel,
K. Berg-Sorensen, L. Oddershede, and R. Metzler, In \textit{Vivo} Anomalous
Diffusion and Weak Ergodicity Breaking of Lipid Granules, Phys. Rev. Lett. 
\textbf{106}, 048103 (2011).

\bibitem{simon} A. V. Weigel, B. Simon, M. M. Tamkun, and D. Krapf, Ergodic
and nonergodic processes coexist in the plasma membrane as observed by
single-molecule tracking, Proc. Natl. Acad. Sci. U.S.A. \textbf{108}, 6438
(2011).

\bibitem{geneston} B. J. West, E. L. Geneston, and P. Grigolini, Maximizing
information exchange between complex networks, Phys. Rep. \textbf{468}, 1
(2008); and references there in.

\bibitem{west} N. Piccinini, D. Lambert, B. J. West, M. Bologna, and P.
Grigolini, Non-ergodic Complexity Management, arXiv:1511.08140 (2016); E.
Geneston, R. Tuladhar, M. T. Beig, M. Bologna, and P. Grigolini, Ergodicity
Breaking and Localization, arXiv:1601.02879 (2015).

\bibitem{turbulence} L. Silvestri, L. Fronzoni, P. Grigolini, and P.
Allegrini, Event-Driven Power-Law Relaxation in Weak Turbulence, Phys. Rev.
Lett. \textbf{102}, 014502 (2009).

\bibitem{ross} S. Bianco, M. Ignaccolo, M. S. Rider, M. J. Ross, P. Winsor,
and P. Grigolini, Brain, Music and non-Poisson Renewal Processes, Phys. Rev.
E \textbf{75}, 061911 (2007).

\bibitem{gunter} G. M. Sch\"{u}tz and S. Trimper, Elephants can always
remember: Exact long-range memory effects in a non-Markovian random walk,
Phys. Rev. E \textbf{70}, 045101(R) (2004).

\bibitem{kim} H. Kim, Anomalous diffusion induced by enhancement of memory,
Phys. Rev. E \textbf{90}, 012103 (2014).

\bibitem{kursten} R. K\"{u}rsten, Random recursive trees and the elephant
random walk, Phys. Rev. E \textbf{93}, 032111 (2016).

\bibitem{gandi} J. C. Cressoni, M. A. A. da Silva, and G. M. Viswanathan,
Amnestically Induced Persistence in Random Walks, Phys. Rev. Lett. \textbf{98%
}, 070603 (2007); A. S. Ferreira, J. C. Cressoni, G. M. Viswanathan, and M.
A. Alves da Silva, Anomalous diffusion in non-Markovian walks having
amnestically induced persistence, Phys. Rev. E \textbf{81}, 011125 (2010);
J. C. Cressoni, G. M. Viswanathan, and M. A. A. da Silva, Exact solution of
an anisotropic 2D random walk model with strong memory correlations, J.
Phys. A \textbf{46}, 505002 (2013).

\bibitem{kenkre} V. M. Kenkre, Analytic formulation, Exact Solutions, and
Generalizations of the elephant and the Alzheimer Random Walks,
arXiv:0708.0034 (2007).

\bibitem{katja} N. Kumar, U. Harbola, and K. Lindenberg, Memory-induced
anomalous dynamics: emergence of diffusion, subdiffusion, and superdiffusion
from a single random walk model, Phys. Rev. E \textbf{82}, 021101 (2010).

\bibitem{boyer} D. Boyer and J. C. Romo-Cruz, Solvable random-walk model
with memory and its relations with Markovian models of anomalous diffusion,
Phys. Rev. E \textbf{90}, 042136 (2014).

\bibitem{esguerra} F. N. C. Paraan and J. P. Esguerra, Exact moments in a
continuous time random walk with complete memory of its history, Phys. Rev.
E \textbf{74}, 032101 (2006).

\bibitem{hanel} R. Hanel and S. Thurner, Generalized (c,d)-Entropy and Aging
Random Walks, Entropy \textbf{15}, 5324 (2013).

\bibitem{feller} W. Feller,\textit{\ An introduction to probability theory
and applications}, Vol. I \& II, (John Wiley \& Sons, 1967).

\bibitem{norman} N. L. Johnson and S. Kotz, \textit{Urn Models and Their
Application}, (John Wiley\&Sons, 1977).

\bibitem{pitman} J. Pitman, \textit{Combinatorial Stochastic Processes},
(Springer 2006).

\bibitem{queen} D. Blackwell and J. B. MacQueen, Fergurson distributions via
P\'{o}lya urn schemes, The Annals of Statistics \textbf{1}, 353 (1973).

\bibitem{budini} A. A. Budini, Central limit theorem for a class of globally
correlated random variables, arXiv:1603.07314.
\end{thebibliography}
\end{document}